\documentstyle[12pt,epsfig,citesort]{article}
\newcommand{\Z}{{\sf Z \!\!\! Z}}

\newcommand{\Psibar}{\overline \Psi}
\newcommand{\psibar}{\overline \psi}
\makeatletter
\@addtoreset{equation}{section}
\makeatother

\title{Dimensional Reduction of Fermions in \\  \vspace*{3mm}
Brane Worlds of the Gross-Neveu Model}

\author{W. Bietenholz$^{{\rm \, a}}$, A. Gfeller$^{{\rm \, b}}$, and 
U.-J. Wiese$^{{\rm \, b}}$\footnote{on leave from MIT}
\\ \\
$^{{\rm a}}$ Institut f\"ur Physik, Humboldt Universit\"at zu Berlin \\
Newtonstrasse 15, D-12489 Berlin, Germany \\ \\
$^{{\rm b}}$ Institut f\"ur Theoretische Physik, Universit\"at Bern \\
Sidlerstrasse 5, CH-3012 Bern, Switzerland \\}

\begin{document} 

\maketitle

\begin{center}
Preprint \ \ HU-EP-03/58, \ SFB-CPP-03-40
\end{center}

\begin{abstract} \normalsize

We study the dimensional reduction of fermions, both in the symme\-tric and in 
the broken phase of the 3-d Gross-Neveu model at large $N$. In particular, in
the broken phase we construct an exact solution for a stable brane world 
consisting of a domain wall and an anti-wall. A left-handed 2-d fermion 
localized on the domain wall and a right-handed fermion localized on the 
anti-wall communicate with each other through the 3-d bulk. In this way they 
are bound together to form a Dirac fermion of mass $m$. As a consequence of 
asymptotic freedom of the 2-d Gross-Neveu model, the 2-d correlation length 
$\xi = 1/m$ increases exponentially with the brane separation. Hence, from the 
low-energy point of view of a 2-d observer, the separation of the branes 
appears very small and the world becomes indistinguishable from a 2-d 
space-time. Our toy model provides a mechanism for brane stabilization: branes 
made of fermions may be stable due to their baryon asymmetry. Ironically, our 
brane world is stable only if it has an extreme baryon asymmetry with all 
states in this ``world'' being completely filled. 

\end{abstract}
 
\maketitle
 
\newpage

\section{Introduction}

Why is gravity so weak? Or equivalently, why are the proton (and other hadrons)
so light compared to the Planck scale? As Wilczek has explained nicely 
\cite{Wil02}, the solution to this hierarchy puzzle results from asymptotic 
freedom. Without any fine-tuning of the bare gauge coupling at distances as 
short as the Planck length, a 4-d non-Abelian gauge theory like QCD produces
a correlation length $\xi$ that is larger than the shortest length scale in the
problem, by a factor exponentially large in the inverse coupling. The inverse 
correlation length defines a mass scale $m = 1/\xi$ --- the proton mass in 
Wilczek's example --- that is hence exponentially smaller than the Planck 
scale. Since we consist mostly of these very light particles (protons and 
neutrons) in natural units of their mass we experience gravity as a very weak 
force.

Lattice gauge theorists benefit from the absence of a hierarchy problem in 
numerical simulations of gauge theories. In this case the shortest length scale
(and hence the analog of the Planck length) is the lattice spacing. Again, 
thanks to asymptotic freedom, without any fine-tuning of the bare gauge 
coupling, the physical correlation length can be made arbitrarily large in 
lattice units. Hence, putting aside practical difficulties due to the 
limited power of computers, there is no problem of principle in approaching the
continuum limit in 4-d Yang-Mills theories on the lattice. Unfortunately, the
situation is not as simple in full lattice QCD including quarks. In fact, for a
long time lattice field theorists have suffered from a hierarchy problem in 
the fermion sector. This problem arises when one removes the unwanted doubler
fermions by breaking chiral symmetry explicitly, for example, by introducing a
Wilson term in the lattice fermion action. Recovering chiral symmetry in the
continuum limit then requires a delicate fine-tuning of the bare fermion mass. 
This is not only a pain in practical numerical computations, but should be
considered a serious problem of principle at the heart of the nonperturbative 
regularization of theories with a chiral symmetry. When one works in the 
continuum, one often takes chiral symmetry for granted because it can be
maintained in dimensional regularization. However, the subtleties related to
the definition of $\gamma_5$ in the framework of dimensional regularization are
just another aspect of the same deep problem of chiral symmetry that is 
obvious on the lattice.

Returning to the hierarchy puzzle, we should ask why the quarks inside the
proton are light. In particular, we should still be puzzled by the fact that we
consist of not just of gluons, but also of light quarks. Certainly, if one 
imagines Wilson's lattice QCD as a (highly simplified) model for the short 
distance physics of our world, without unnatural fine-tuning, quarks would live
at the lattice spacing scale, while gluons (or, more precisely, glueballs) are 
naturally light. Fortunately, the long-standing hierarchy problem of lattice 
fermions --- and hence of the nonperturbative regularization of chiral symmetry
--- has recently been solved. Based on work of Callan and Harvey \cite{Cal85}, 
the first step in this direction was taken by Kaplan \cite{Kap92} who realized 
that massless lattice fermions arise naturally, i.e.\ without fine-tuning, as 
zero-modes localized on a domain wall embedded in a higher-dimensional 
space-time. For example, left-handed fermions in four dimensions can be 
localized on a domain wall that represents a 3-brane embedded in a 5-d 
space-time. Similarly, right-handed fermions can be localized on an anti-wall. 
By keeping the wall and the anti-wall at a sufficiently large distance, i.e.\ 
by spatially separating left- and right-handed fermions in the extra dimension,
they are prevented from mixing strongly with one another. Thus, they are 
protected from picking up a large mass and are naturally light. Since 
$\gamma_5$ appears in its action, the 5-d theory itself does not even have a 
chiral symmetry. Hence, in contrast to four dimensions, a Wilson term in the 
5-d lattice action removes the doubler fermions without doing damage to
chiral symmetry. Overlap lattice fermions \cite{Neu93} also 
live in an extra dimension and are closely related to domain wall fermions. 
When one integrates out the extra dimension, the 4-d version of overlap 
fermions satisfies the so-called Ginsparg-Wilson relation \cite{Gin82}. As 
L\"uscher first realized, this relation implies a lattice version of chiral 
symmetry \cite{Lue98} which led him to a spectacular breakthrough: the 
nonperturbative construction of lattice chiral gauge theories \cite{Lue99}. 
Perfect and classically perfect lattice fermion actions 
\cite{Wie93,Bie96,DeG97} also obey the Ginsparg-Wilson relation \cite{Has98}. 
However, the explicit construction of such actions is a delicate problem that 
can itself be considered a form of fine-tuning. Fermion actions that naturally 
obey the Ginsparg-Wilson relation without fine-tuning, on the other hand, can 
be related to the physics in a higher-dimensional space-time. The existence of 
light 4-d fermions may hence be a hint to the physical reality of extra 
dimensions. Indeed, brane worlds embedded in a higher-dimensional space-time 
provide a very interesting perspective on ordinary 4-d physics 
\cite{Ark98,Ran99}.

In present lattice QCD applications of domain wall or overlap fermions the 
extra dimension is not taken physically seriously. In particular, the gluon 
field is usually frozen in the fifth dimension, thus violating locality in the
extra dimension. For left- and right-handed domain wall fermions localized on a
wall and an anti-wall, one takes the chiral limit by separating the wall from 
the anti-wall in the extra dimension, while the gluonic correlation length is
kept fixed. Consequently, one approaches the chiral limit before the continuum 
limit, and the extra dimension does not have a physically meaningful extent. 
In order to cancel singularities resulting from the 5-d bulk, one introduces 
ghost fields violating the spin-statistics theorem. While all this is not wrong
if one only wants to construct 4-d QCD, it is unnatural if one takes the fifth 
dimension seriously. 

There is an alternative nonperturbative formulation of QCD (and other field 
theories) --- D-theory \cite{Cha97} --- 
in which the extra dimension is used in 
a more natural way. In D-theory the familiar classical fields emerge 
dynamically from discrete variables (such as quantum spins or quantum links)
which undergo dimensional reduction. In this formulation of QCD, gluons emerge 
dynamically from higher-dimensional physics. Starting with a 5-d quantum link
model from which a non-Abelian Coulomb phase with massless gluons emerges, 
compactification of the fifth dimension leads to a correlation length (i.e.\ an
inverse glueball mass) that is exponentially large in the size of the extra 
dimension. As a consequence of asymptotic freedom of 4-d QCD, the size of the 
extra dimension then shrinks to zero in physical units as one approaches the 
continuum limit. Domain wall fermions fit naturally into this framework. In 
particular, the chiral and continuum limits are reached simultaneously, the 
theory is fully local in the extra dimension, no unphysical ghost fields are 
needed, and the extra dimension completely disappears in the continuum limit 
via dimensional reduction. If one imagines the D-theory regularization of QCD 
as an (again highly simplified) short distance model of our world, not just
gluons, but also quarks are naturally light. This solves the second part of the
hierarchy puzzle: if we assume the existence of a fifth dimension, we need no 
longer be puzzled why we also consist of light quarks.

In this paper we use the 3-d Gross-Neveu model \cite{Gro74} at large $N$ as an 
analytically soluble toy model for illustrating the issues discussed above. 
Using the standard procedure, domain wall fermions were applied to the 
Gross-Neveu model in Ref.\ \cite{Izu00}. 
In that calculation the fermionic left- and 
right-handed zero-modes localized on a wall and an anti-wall are coupled by 
hand. While this is sufficient if one just wants to construct the 2-d 
Gross-Neveu model, it violates locality in the extra dimension. This is 
unnatural if one takes that dimension seriously. In this paper, in the spirit 
of D-theory, we do take the extra dimension physically seriously and let the 
left- and right-handed zero-modes communicate through the bulk of the extra 
dimension by a locally implemented 4-fermion interaction. Then, just as in QCD,
the D-theory mechanism of dimensional reduction leads to naturally light 
fermions without fine-tuning. In our calculation not even the domain walls on 
which fermion states can be localized are put by hand. They arise dynamically 
from the spontaneous breakdown of the $\Z(2)$ chiral symmetry of the 
Gross-Neveu model. The walls play the role of branes and hence our construction
can be viewed as an attempt to construct stable brane worlds. We find exact 
brane world solutions just relying on the dynamics of the underlying 
higher-dimensional theory, without making ad-hoc assumptions about where the 
branes shall be located. In realistic brane world constructions stabilizing the
branes is a highly non-trivial issue. In our model we construct an exact 
solution for a fully consistent and stable wall-anti-wall configuration. 
Although our toy model is very simple, it reveals an interesting mechanism for 
brane stabilization: if branes are made of fermions (and thus have a baryon 
asymmetry) baryon number conservation in the higher-dimensional theory may 
imply brane stability. Unfortunately, in our toy ``world'' the baryon asymmetry
is so large that all states are occupied with fermions and any potentially 
non-trivial physics is completely Pauli-blocked.

In Section 2 basic features of the 2-d Gross-Neveu model are reviewed, while 
Section 3 discusses the 3-d Gross-Neveu model. In Section 4 the 3-d Gross-Neveu
model is considered in the chirally symmetric phase. After compactification of 
the third dimension, the model undergoes dimensional reduction to the 2-d 
Gross-Neveu model by the generation of a correlation length that is 
exponentially large in the size of the extra dimension. Section 5 discusses 
dimensional reduction from the chirally broken phase of the 3-d model, using 
configurations with either one wall or a wall-anti-wall pair. In the 
wall-anti-wall case a naturally light Dirac fermion is generated with its left-
and right-handed components being localized on the two walls. Again, the theory
undergoes dimensional reduction to the 2-d Gross-Neveu model. The emerging 
chiral order parameter of the 2-d theory as well as the brane tensions and the 
issue of brane world stability are also discussed. Finally, Section 6 contains 
our conclusions. A synopsis of this work is given in Ref.\ \cite{Lat03}.

\section{The 2-d Gross-Neveu Model}

In this Section, we introduce the 2-d Gross-Neveu model \cite{Gro74} as the 
target theory which we will later obtain from dimensional reduction of the 
corresponding 3-d model. Various 2-d field theories, including the Gross-Neveu 
model, were recently reviewed in \cite{Sch01}. We consider the Gross-Neveu 
model at large $N$ in the continuum. Its Euclidean action is given by
\begin{equation}
\label{action2}
S[\psibar,\psi] = \int d^2x \left[\psibar \gamma_{\mu} \partial_{\mu} \psi - 
\frac{g}{2N} \left(\psibar \psi \right)^2\right].
\end{equation}
We have suppressed a flavor index that runs from 1 to $N$ and gives rise to a
global $U(N)$ flavor symmetry. The parameter $g$ is the dimensionless 4-fermion
coupling which remains fixed in the 't Hooft limit $N \rightarrow \infty$. The 
explicit factor $1/N$ ensures that the interaction term stays of the same
magnitude as the kinetic term in this limit. In addition to the $U(N)$ flavor 
symmetry, the model has a $\Z(2)$ chiral symmetry
\begin{eqnarray}
\label{symmetry2}
&&\psi_L(x)' = \pm \psi_L(x), \
\psibar_L(x)' = \pm \psibar_L(x), \nonumber \\
&&\psi_R(x)' = \mp \psi_R(x), \
\psibar_R(x)' = \mp \psibar_R(x),
\end{eqnarray}
where
\begin{equation}
\psi_{R,L}(x) = \frac{1\pm \gamma_3}{2} \psi(x), \
\psibar_{R,L}(x) = \psibar(x) \frac{1\mp \gamma_3}{2}.
\end{equation}

Let us consider the $N = \infty$ limit and derive a gap equation that describes
the spontaneous breakdown of the $\Z(2)$ symmetry of eq.(\ref{symmetry2}). 
First, we introduce an auxiliary scalar field
\begin{equation}
\phi(x) = \frac{g}{N} \psibar(x) \psi(x),
\end{equation}
representing the chiral order parameter $\psibar \psi$. It linearizes the 
4-fermion interaction, such that
\begin{equation}
S[\psibar,\psi,\phi] = \int d^2x 
\left[\psibar \gamma_{\mu} \partial_{\mu} \psi 
- \psibar \psi \phi + \frac{N}{2 g} \phi^2 \right].
\end{equation}
The partition function then takes the form
\begin{equation}
Z = \int {\cal D}\psibar {\cal D}\psi \exp(- S[\psibar,\psi]) =
\int {\cal D}\psibar {\cal D}\psi {\cal D}\phi \exp(- S[\psibar,\psi,\phi]).
\end{equation}
In the large $N$ limit we may restrict $\phi(x)$ to a space-time-independent 
constant $\phi$ \cite{Gro74}. Doing so, the action in momentum space takes the 
form
\begin{equation}
S[\psibar,\psi,\phi] = \frac{1}{(2\pi)^2} \int d^2k \
\psibar(- k)[i \gamma_\mu k_\mu - \phi] \psi(k) + 
\frac{N V}{2 g} \phi^2,
\end{equation}
where $V$ is the volume of space-time. We integrate out the fermion fields
\begin{equation}
\int {\cal D}\psibar {\cal D}\psi \ \exp(- S[\psibar,\psi,\phi]) = 
\exp(- N V V_{eff}(\phi)),
\end{equation}
in order to obtain the effective potential $V_{eff}(\phi)$ of the chiral order 
parameter. In the infinite volume limit the fermion integration yields
\begin{equation}
\label{epot2}
V_{eff}(\phi) = - \frac{1}{(2 \pi)^2} \int d^2k \
\log\left(k^2 + \phi^2\right) + \frac{1}{2 g} \phi^2.
\end{equation}
Using eq.(\ref{epot2}), the minimum $\phi_0$ of the effective potential is 
determined by
\begin{equation}
\frac{dV_{eff}}{d\phi}(\phi_0) = - \frac{1}{(2\pi)^2} \int d^2k \ 
\frac{2 \phi_0}{k^2 + \phi_0^2} + \frac{\phi_0}{g} = 0,
\end{equation}
which implies the gap equation
\begin{equation}
\frac{1}{(2\pi)^2} \int d^2k \ \frac{2}{k^2 + \phi_0^2} = \frac{1}{g}.
\end{equation}
In order to solve the gap equation, a cut-off $\Lambda_2$ is introduced in 2-d 
momentum space. The ultraviolet limit $\Lambda_2 \gg \phi_0$ is reached for 
$g \ll 1$ and one then obtains
\begin{equation}
\label{massgap2}
\phi_0 = m = \Lambda_2 \exp(- \frac{\pi}{g}).
\end{equation}
The exponential factor in eq.(\ref{massgap2}) is a manifestation of asymptotic 
freedom of the 2-d Gross-Neveu model. The factor $\pi$ in the exponent is the
inverse of the corresponding 1-loop $\beta$-function coefficient. Due to 
spontaneous chiral symmetry breaking, the fermions pick up a mass $m = \phi_0$.
The cut-off $\Lambda_2$ is removed by varying the bare coupling $g$ such that 
the physical fermion mass $m$ remains fixed. After removing the cut-off, the 
effective potential takes the form
\begin{equation}
V_{eff}(\phi) = \frac{\phi^2}{4 \pi}(\log\frac{\phi^2}{\phi_0^2} - 1),
\end{equation}
which is depicted in Figure 1.
\begin{figure}[htb]
\begin{center}
\epsfig{figure=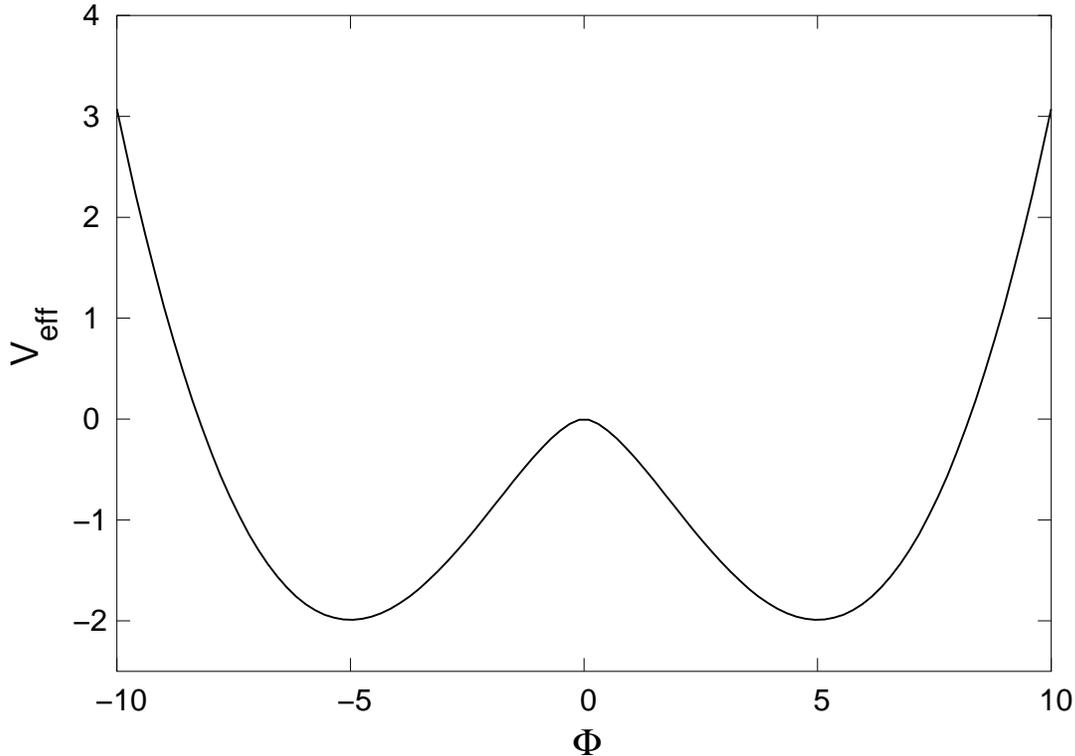,width=10cm,angle=270}
\end{center}
\caption{\it The effective potential for $d = 2$ (with $\phi_0 = 5$).}
\end{figure}

\section{The 3-d Gross-Neveu Model}

Next we consider the 3-d Gross-Neveu model \cite{Gat92,Ana98,Hof02} with the 
Euclidean action
\begin{equation}
\label{action3}
S[\Psibar,\Psi] = \int d^3x \left[\Psibar \gamma_\mu \partial_\mu \Psi +
\Psibar \gamma_3 \partial_3 \Psi - 
\frac{G}{2N} \left(\Psibar \Psi \right)^2\right].
\end{equation}
In three dimensions the 4-fermion coupling $G$ has the dimension of a 
length. We will soon compactify the third direction or endow it with domain 
walls and obtain the 2-d Gross-Neveu model by means of dimensional reduction. 
Hence, in this case, the 3-direction is not Euclidean time but just an 
additional spatial dimension which will ultimately become invisible. Instead, 
we choose the 2-direction to represent Euclidean time. In three dimensions 
there is no chiral symmetry because $\gamma_3$ appears explicitly in the 
action. Still, the 3-d action has a $\Z(2)$ symmetry which reduces to the 
chiral symmetry of the 2-d Gross-Neveu model after dimensional reduction, but 
which also involves a spatial inversion in the 3-direction,
\begin{eqnarray}
\label{symmetry3}
\!\!\!\!\!&&\Psi_L(x_1,x_2,x_3)' = \pm \Psi_L(x_1,x_2,- x_3), \
\Psibar_L(x_1,x_2,x_3)' = \pm \Psibar_L(x_1,x_2,- x_3), \nonumber \\
\!\!\!\!\!&&\Psi_R(x_1,x_2,x_3)' = \mp \Psi_R(x_1,x_2,- x_3), \
\Psibar_R(x_1,x_2,x_3)' = \mp \Psibar_R(x_1,x_2,- x_3).
\end{eqnarray}
Later we will consider the $x_3$-independent zero-mode that determines the 
physics of the dimensionally reduced 2-d Gross-Neveu model. Then 
eq.(\ref{symmetry3}) reduces to the usual 2-d $\Z(2)$ chiral symmetry of 
eq.(\ref{symmetry2}).

As in the 2-d case, we take the large $N$ limit and derive a gap equation
\begin{equation}
\frac{1}{(2\pi)^3} \int d^3k \ \frac{2}{k^2 + \Phi_0^2} = \frac{1}{G}.
\end{equation}
Introducing a cut-off $\Lambda_3$ in 3-d momentum space, and assuming
$\Phi_0 \neq 0$, in the ultraviolet limit $\Lambda_3 \gg \Phi_0$ we obtain
\begin{equation}
\label{Phi0}
\frac{1}{\pi^2}(\Lambda_3 - \frac{\pi}{2} \Phi_0) = \frac{1}{G} \
\Rightarrow \ \Phi_0 = 2 \pi \left(\frac{1}{G_c} - \frac{1}{G}\right).
\end{equation}
We have introduced the critical coupling constant $G_c = \pi^2/\Lambda_3$.
At strong coupling, $G > G_c$, one has $\Phi_0 > 0$, so we are in the broken
phase. For weak coupling, $G < G_c$, on the other hand, we are in the symmetric
phase. For large $\Lambda_3$ (and up to a trivial additive constant) the 
effective potential reduces to
\begin{equation}
V_{eff}(\Phi) = \frac{1}{6 \pi} |\Phi|^3 - 
\frac{1}{2} \left(\frac{1}{G_c} - \frac{1}{G}\right) \Phi^2.
\end{equation}
In the symmetric phase the effective potential $V_{eff}(\Phi)$ has a single 
minimum at $\Phi = 0$, while in the broken phase it has two degenerate minima
at $\Phi = \pm \Phi_0$. The effective potential in the symmetric and broken
phases is depicted in Figures 2 and 3, respectively.
\begin{figure}[htb]
\begin{center}
\epsfig{figure=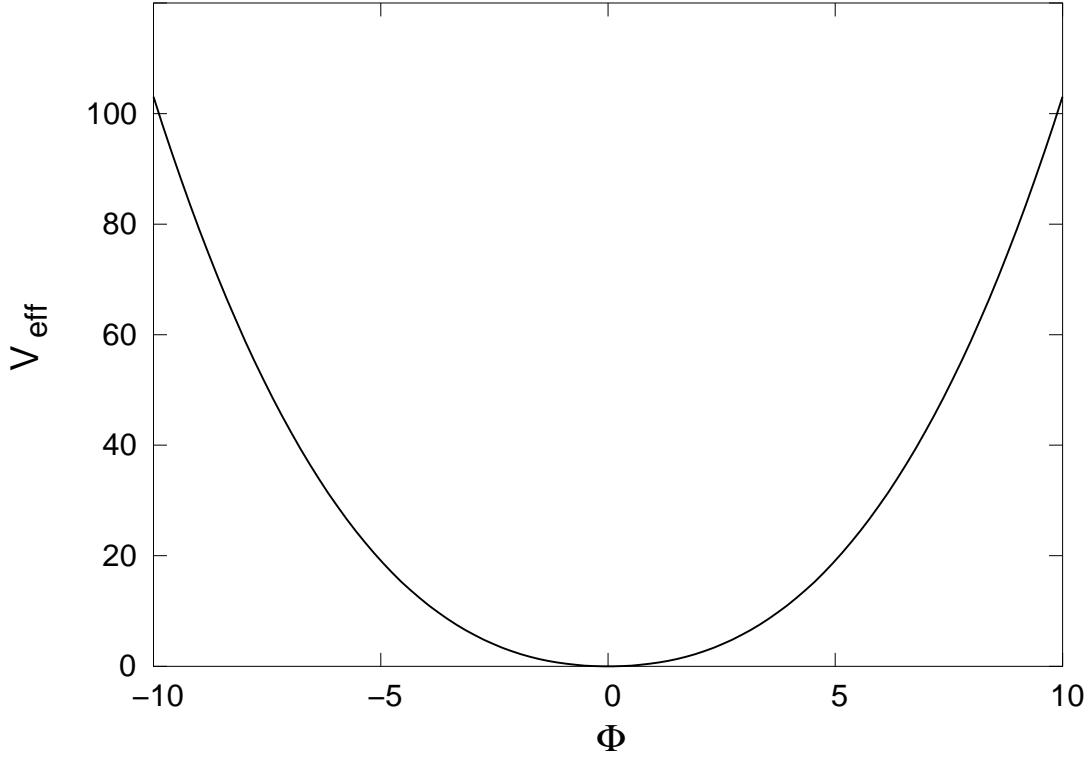,width=10cm,angle=270}
\end{center}
\caption{\it The effective potential for $d = 3$ in the symmetric phase (with
$G_c = 1$ and $G = 0.5$).}
\end{figure}
\begin{figure}[htb]
\begin{center}
\epsfig{figure=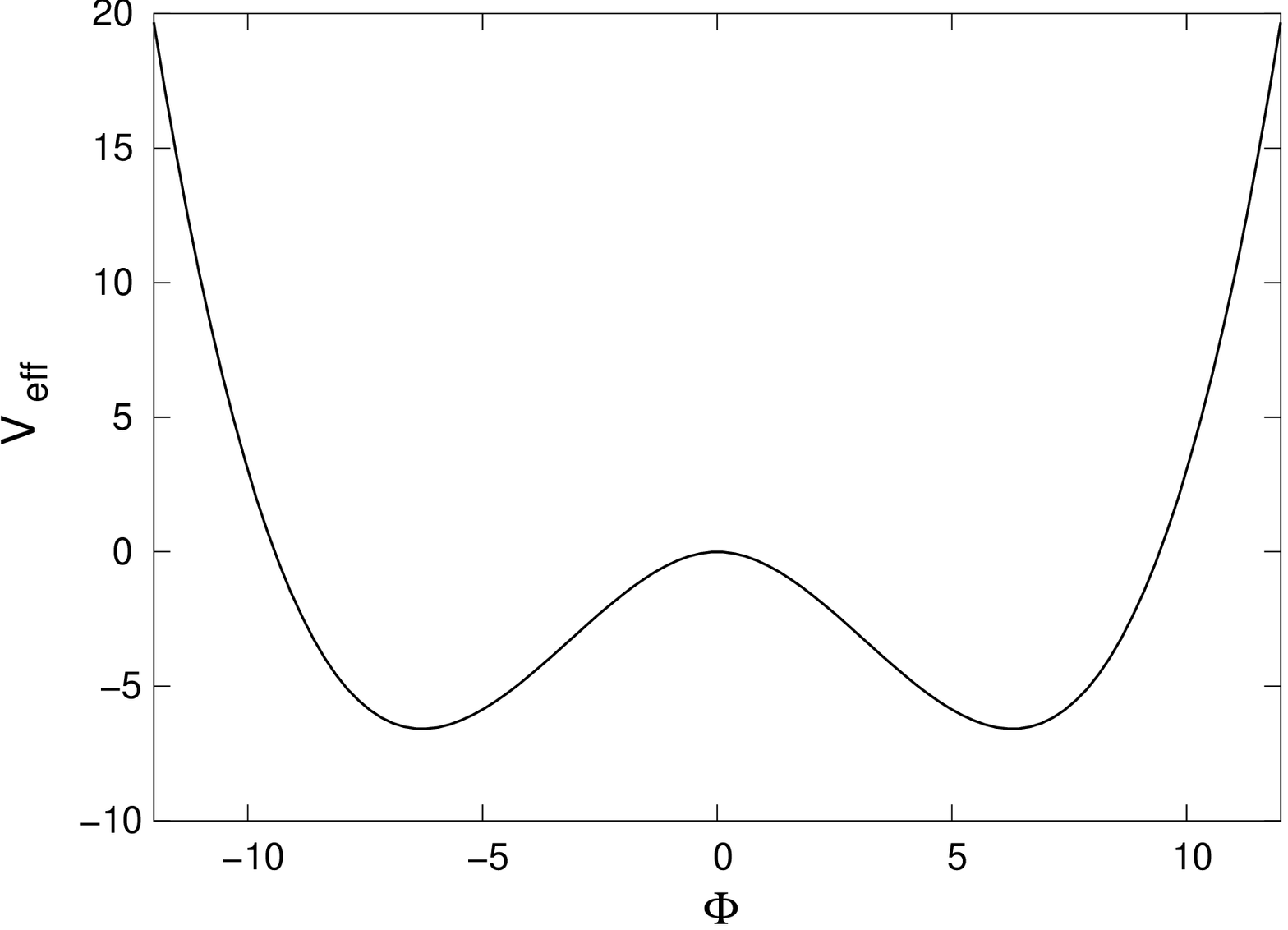,width=10cm,angle=270}
\end{center}
\caption{\it The effective potential for $d = 3$ in the broken phase (with
$G_c = 1$ and $G = 2$).}
\end{figure}

\section{Dimensional Reduction from the Chirally Symmetric Phase}

In this Section we start from the symmetric phase of the 3-d Gross-Neveu model.
By compactifying the third dimension to a circle of circumference $\beta$, the
system is dimensionally reduced to the 2-d Gross-Neveu model. Interestingly, 
the massless fermion that exists in the bulk of the 3-d Gross-Neveu model 
cannot remain massless after compactification. If it would, the size $\beta$ of
the extra dimension would be negligible compared to the then infinite fermionic
correlation length, and the model would become a massless 2-d Gross-Neveu 
model. However, as we have seen, in the 2-d Gross-Neveu model the $\Z(2)$ 
chiral symmetry is spontaneously broken for all values of the coupling constant
$g$. Hence, as the third dimension is compactified, the previously massless 3-d
fermion must necessarily pick up a mass. Consequently, its correlation length 
$\xi = 1/m$ will become finite. Then the question arises if $\xi$ is large or 
small compared to $\beta$. 

Let us study the mechanism of dimensional reduction from the symmetric phase in
some detail. The 3-d gap equation with periodic boundary conditions in the 
third direction takes the form
\begin{equation}
\frac{1}{(2\pi)^3} \int d^2k \ \frac{2 \pi}{\beta} \sum_{n \in \Z} 
\frac{2}{k_1^2 + k_2^2 + (2 \pi n/\beta)^2 + \Phi_0^2} = \frac{1}{G}.
\end{equation}
To evaluate the sum, we use the Poisson resummation formula and we obtain
\begin{equation}
\sum_{n \in \Z} \frac{2}{k_1^2 + k_2^2 + (2 \pi n/\beta)^2 + \Phi_0^2} =
\frac{\beta \coth\left(\beta \sqrt{k_1^2 + k_2^2 + \Phi_0^2}/2\right)}
{\sqrt{k_1^2 + k_2^2 + \Phi_0^2}}.
\end{equation}
The gap equation then reads
\begin{equation}
\frac{1}{(2\pi)^2} \int d^2k \
\frac{\coth\left(\beta \sqrt{k_1^2 + k_2^2 + \Phi_0^2}/2\right)}
{\sqrt{k_1^2 + k_2^2 + \Phi_0^2}} = \frac{1}{G}.
\end{equation}
Again, we introduce the cut-off $\Lambda_2$ in 2-d momentum space and perform 
the integral. For large $\Lambda_2$ we obtain
\begin{equation}
\label{massgap3to2}
\sinh\frac{\beta \Phi_0}{2} = 
\exp(\pi \beta \left(\frac{\Lambda_2}{2 \pi} - \frac{1}{G}\right)).
\end{equation}
In order to match the 2-d cut-off $\Lambda_2$ to the 3-d cut-off $\Lambda_3$, 
for a moment we consider the broken phase of the 3-d model. Then $\Phi_0$ 
approaches its constant bulk value as $\beta \rightarrow \infty$, such that
eq.(\ref{massgap3to2}) implies
\begin{equation}
\label{cutoff}
\Phi_0 = 2 \pi \left(\frac{\Lambda_2}{2 \pi} - \frac{1}{G}\right).
\end{equation}
We match this with the previous result of eq.(\ref{Phi0}) by identifying
\begin{equation}
\frac{\Lambda_2}{2 \pi} = \frac{1}{G_c} = \frac{\Lambda_3}{\pi^2}.
\end{equation}
Returning to the symmetric phase of the 3-d model, $\Phi_0$ vanishes as 
$\beta \rightarrow \infty$ and eq.(\ref{massgap3to2}) implies
\begin{equation}
\label{xi}
\xi = \frac{1}{m} = \frac{1}{\Phi_0} = \frac{\beta}{2} 
\exp(\pi \beta \left(\frac{1}{G} - \frac{1}{G_c}\right)).
\end{equation}
Interestingly, the correlation length $\xi$ becomes exponentially larger than 
$\beta$ as $\beta$ itself goes to infinity. Hence, counter-intuitively, as the 
size of the third dimension becomes large (in units of the inverse cut-off
$1/\Lambda_3$) the physical correlation length of the fermion increases 
exponentially and the low-energy physics reduces to the one of the 2-d 
Gross-Neveu model. There is a hierarchy of three separate length scales in this
problem. The shortest length scale is determined by the inverse cut-off 
$1/\Lambda_3$. The next scale $\beta \gg 1/\Lambda_3$ is the size of the extra 
dimension. Finally, the largest length scale $\xi \gg \beta$ is dynamically 
generated by spontaneous chiral symmetry breaking in the dimensionally reduced 
2-d model. In fact, the size $\beta$ of the extra dimension plays the role of 
the inverse cut-off for the 2-d Gross-Neveu model that arises by means of 
dimensional reduction. Identifying the coupling constant of the dimensionally 
reduced model as
\begin{equation}
\frac{1}{g} = \beta \left(\frac{1}{G} - \frac{1}{G_c}\right),
\end{equation}
eq.(\ref{xi}) turns into
\begin{equation}
\label{mass3to2}
m = \frac{1}{\xi} = \frac{2}{\beta} \exp(- \frac{\pi}{g}),
\end{equation}
which is nothing but the asymptotic freedom relation eq.(\ref{massgap2}) of the
2-d Gross-Neveu model. In particular, we identify $2/\beta$ as the effective
cut-off of the dimensionally reduced model. The same subtle mechanism of 
dimensional reduction was first observed in the 3-d $O(3)$ sigma model 
\cite{Cha88,Has91}. However, in that case dimensional reduction results only 
when one starts in the broken phase of the 3-d model, which contains massless 
Goldstone bosons. As a consequence of the Mermin-Wagner-Coleman Theorem, these 
particles cannot remain massless as the theory dimensionally reduces to two 
dimensions. Since the $\Z(2)$ symmetry of the Gross-Neveu model is discrete, 
the Theorem does not apply in this case. In the 3-d Gross-Neveu model the 
massless phase is the one without spontaneous symmetry breaking. Still, both 
the 2-d $O(3)$ model and the 2-d Gross-Neveu model are massive, although for 
different reasons. In the next Section we will see how dimensional reduction 
arises starting from the broken phase of the 3-d Gross-Neveu model. Figure 4 
shows the chiral condensate (or equivalently the mass gap) as a function of $G$
both for finite and for infinite $\beta$.
\begin{figure}[htb]
\begin{center}
\epsfig{figure=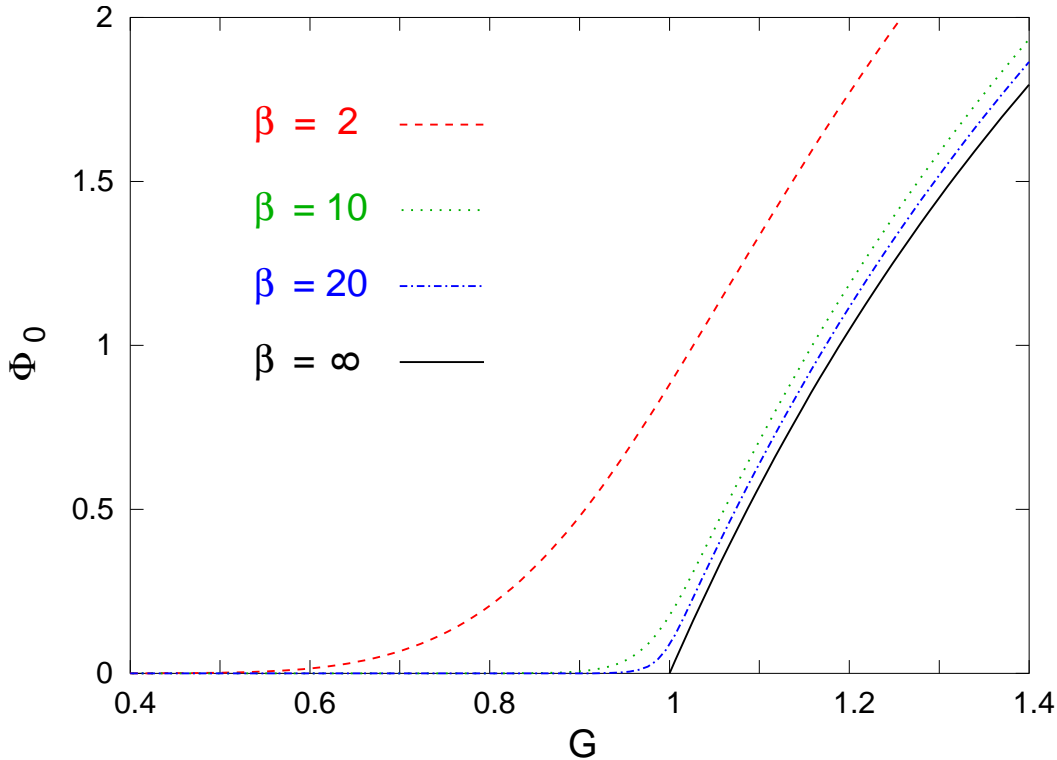,width=10cm,angle=270}
\end{center}
\caption{\it The mass gap for $d = 3$ as a function of $G$ at various values of
$\beta$ (with $G_c = 1$).} 
\end{figure}

From the point of view presented in the Introduction, the way we have obtained
the 2-d Gross-Neveu model by means of dimensional reduction from the symmetric 
phase of the 3-d model is not really satisfactory. Since we studied the large 
$N$ limit, an analytic calculation was possible in the continuum using a 
momentum cut-off that does not break chiral symmetry explicitly. On the other 
hand, a fully nonperturbative treatment at finite $N$ would require a 
formulation on the lattice, which leads to the fermion doubling problem. For 
even $N$ this problem can be avoided using staggered fermions. However, for odd
$N$ one would use Wilson fermions \footnote{Taking the square-root of the 
staggered fermion determinant is unsatisfactory, unless one can show that this 
procedure does not lead to a violation of locality in the continuum limit.} and
thus break chiral symmetry explicitly by the regulator. Recovering chiral 
symmetry in the continuum limit then requires unnatural fine-tuning of the bare
fermion mass. The idea of domain wall fermions is to construct naturally light 
fermions via dimensional reduction from a higher dimension. In the next 
Section, we will see explicitly how this works in the broken phase of the 3-d 
Gross-Neveu model. Dimensional reduction from the symmetric phase, on the other
hand, requires a 3-d massless fermion to begin with. In a fully nonperturbative
lattice calculation at finite $N$ this would require unnatural fine-tuning at 
the level of the 3-d theory. Hence, the above mechanism of dimensional 
reduction from the symmetric phase does not work naturally (i.e.\ without 
fine-tuning) when the cut-off violates chiral symmetry. This is why we now turn
to the chirally broken phase.

\section{Dimensional Reduction from the Chirally Broken Phase}

In this Section we describe how the 2-d Gross-Neveu model can arise via
dimensional reduction from the broken phase of the 3-d model. If one would
again compactify the third dimension to a circle, the finite correlation length
$\xi = 1/m = 1/\Phi_0$ of the fermion in the 3-d bulk would never become
large compared to $\beta$. Hence, with periodic boundary conditions, 
dimensional reduction would not happen from the broken phase, even in the
$\beta \rightarrow 0$ limit, because the ratio $\beta/\xi = \beta \Phi_0
\rightarrow 2 \log(1 + \sqrt{2})$ remains non-zero in this limit. 

Light fermions in $2n$ dimensions can arise naturally by dimensional reduction
from a massive theory in $(2n+1)$ dimensions as zero-modes localized on a 
domain wall \cite{Kap92}. Interestingly, domain walls indeed exist as stable 
topological objects in the broken phase of the 3-d Gross-Neveu model. We will 
see that a system with a single domain wall has a free massless left-handed 2-d
fermion living on the wall. A system with a stable wall-anti-wall pair 
separated by a distance $\beta$, on the other hand, supports a left-handed 
fermion living on the wall as well as a right-handed fermion living on the 
anti-wall. The left- and right-handed fermionic zero-modes communicate with 
each other through the 3-d bulk between the wall and the anti-wall and pick up 
a mass $m$. Interestingly, in analogy to the mechanism of dimensional reduction
discussed before, the corresponding correlation length $\xi = 1/m$ grows 
exponentially with $\beta$ when one separates the wall from the anti-wall. 
Hence, in units of the 2-d fermion correlation length $\xi$, the size $\beta$ 
of the bulk between the walls shrinks to zero. Consequently, from the 
low-energy point of view of a 2-d observer living on the walls, the 
wall-anti-wall system --- although truly 3-dimensional --- looks exactly 
like a 2-d space-time.

\subsection{A Brane World with a Single Domain Wall}

First we consider a system with a single domain wall separating two distinct
broken phases with chiral condensate values $\Phi_0$ and $- \Phi_0$. The domain
wall itself is described by a field configuration $\Phi(x_3)$ that interpolates
between the two vacuum states, $\Phi(\pm \infty) = \pm \Phi_0$. Determining the
shape $\Phi(x_3)$ of the domain wall is a non-trivial problem. Fortunately, a 
similar problem has been solved a long time ago by Dashen, Hasslacher, and 
Neveu \cite{Das75} in the 2-d Gross-Neveu model. In that case, the topological 
object is not a domain wall but simply a solitonic particle. As we will see, 
the 2-d soliton solution of Ref.\ \cite{Das75} 
naturally extends to three dimensions.
In order to determine $\Phi(x_3)$ we apply the following strategy. First, we 
make an educated guess (the 2-d soliton solution) for $\Phi(x_3)$. Second, we 
integrate out the fermions in the given non-trivial background, and we verify 
explicitly that the resulting $\Psibar \Psi$ self-consistently reproduces 
$\Phi(x_3)$. Our approach is a 3-d generalization of work by Pausch, Thies, and
Dolman \cite{Pau91} and by Feinberg \cite{Fei95}, and is also related to work 
by Chandrasekharan \cite{Cha94}.

The ansatz for the domain wall profile which is obvious from what we know about
2-d solitons is given by the standard kink profile
\begin{equation}
\label{profile}
\Phi(x_3) = \Phi_0 \tanh(\Phi_0 x_3).
\end{equation}
We have arbitrarily centered the domain wall at $x_3 = 0$. Of course, there is 
a trivial translational zero-mode which does not concern us here. In Minkowski 
space-time the fermions propagating in the non-trivial domain wall background 
field $\Phi(x_3)$ are described by the single-particle Dirac Hamiltonian
\begin{equation}
H = \gamma_2[\gamma_1 \partial_1 + \gamma_3 \partial_3 - \Phi(x_3)].
\end{equation}
We choose the Euclidean Dirac matrices to coincide with the Pauli matrices,
$\gamma_i = \sigma_i$. After dimensional reduction the 1-direction will remain 
and --- from the low-energy point of view of a 2-d observer --- the spatial
3-direction becomes invisible. Using translation invariance of the Dirac
equation in both, the spatial 1-direction and in time, we make the separation
ansatz
\begin{equation}
\Psi(x_1,x_3,t) = \Psi(x_3) \exp(i k_1 x_1) \exp(- i E t), 
\end{equation}
which implies
\begin{equation}
\left(\begin{array}{cc} k_1 & i \partial_3 + i \Phi(x_3) \\ 
i \partial_3 - i \Phi(x_3) & - k_1 \end{array} \right)
\left(\begin{array}{c} \Psi_R(x_3) \\ \Psi_L(x_3) \end{array}\right) = 
E \left(\begin{array}{c} \Psi_R(x_3) \\ \Psi_L(x_3) \end{array}\right).
\end{equation}
There is a localized state 
\begin{equation}
\Psi_0(x_3) = \sqrt{\frac{\Phi_0}{2}}
\left(\begin{array}{c} 0 \\ \cosh^{-1}(\Phi_0 x_3) \end{array} \right),
\end{equation} 
with the energy eigenvalue $E_0 = - k_1 > 0$. This state describes a free 
massless left-handed fermion with spatial momentum $k_1$ localized on the 
domain wall. Note that the localized states of positive energy are left-movers 
with $k_1 < 0$. Since in 2-d there is no spin, a left-handed particle is 
indeed simply moving to the left. Similarly, the particles localized on an
anti-wall with the profile $\Phi(x_3) = - \Phi_0 \tanh(\Phi_0 x_3)$ are
right-movers. The wall profile as well as the wave function of the localized
state are illustrated in Figure 5.
\begin{figure}[htb]
\begin{center}
\epsfig{figure=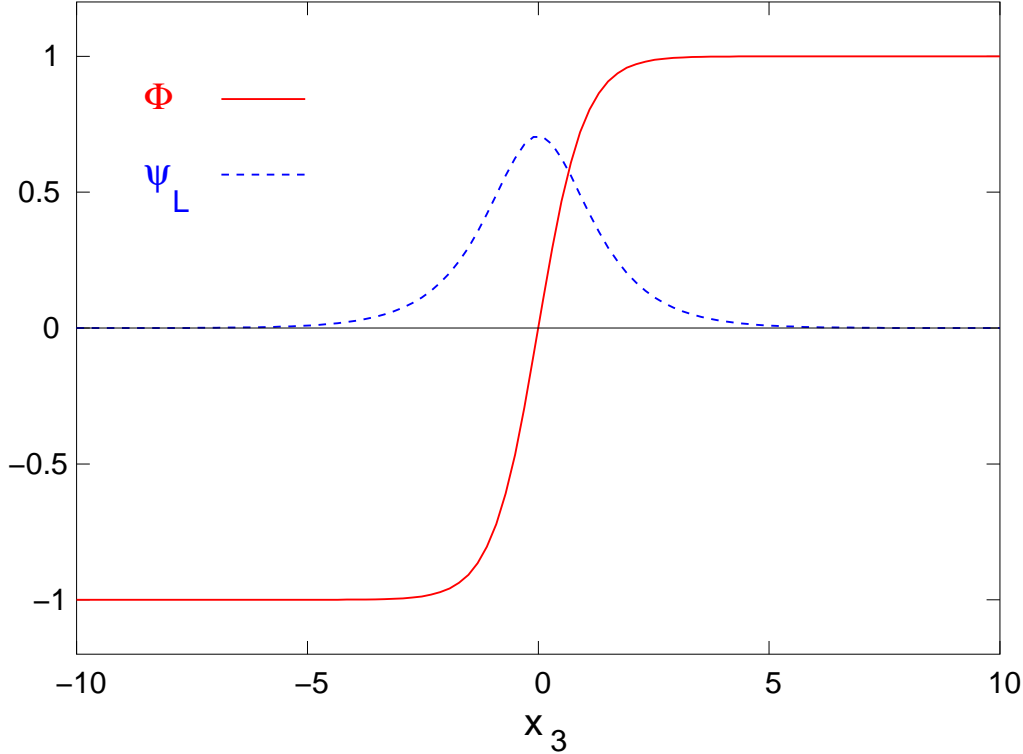,width=10cm,angle=270}
\end{center}
\caption{\it The profile of a single domain wall and the wave function of the
left-handed state $\Psi_0$ localized on this wall (with $\Phi_0 = 1$).}
\end{figure}

The spectrum of the Dirac equation also includes a continuum of 
bulk states
\begin{equation}
\Psi_{k_3}(x_3) = \frac{\exp(i k_3 x_3)}{\sqrt{2 E(E + k_1)}}
\left(\begin{array}{c} i (E + k_1) \\
\Phi_0 \tanh(\Phi_0 x_3) - i k_3 \end{array}\right),
\end{equation}
with the energy eigenvalues $E = \pm \sqrt{k_1^2 + k_3^2 + \Phi_0^2}$. The 
localized and bulk states together form a complete orthonormal basis of the
single particle Hilbert space.

In the next step, we verify explicitly that the solutions from above 
self-con\-sis\-tent\-ly reproduce the domain wall profile of 
eq.(\ref{profile}). For this purpose we evaluate
\begin{eqnarray}
&&\Psibar_0(x_3) \Psi_0(x_3) = \Psi^\dagger_0(x_3) \gamma_2 \Psi_0(x_3) = 0, 
\nonumber \\
&&\Psibar_{k_3}(x_3) \Psi_{k_3}(x_3) = - \frac{1}{E} \Phi_0 \tanh(\Phi_0 x_3) =
- \frac{1}{E} \Phi(x_3).
\end{eqnarray}
Summing over the filled Dirac sea of negative energy states, in the limit 
$\Lambda_2 \gg \Phi_0$ one finds  
\begin{equation}
\frac{G}{N} \Psibar(x_3) \Psi(x_3) = \Phi(x_3) \frac{G}{2 \pi} 
\int_0^{\Lambda_2} \frac{k \, dk}{\sqrt{k^2 + \Phi_0^2}} = \Phi(x_3) 
\frac{G}{2 \pi}(\Lambda_2 - \Phi_0) = \Phi(x_3),
\end{equation}
i.e.\ the ansatz of eq.(\ref{profile}) for the profile of the domain wall is
indeed reproduced by the actual $\Psibar \Psi$ of the fermions living in the
corresponding background field. The factor $1/N$ has canceled because each 
negative energy state is filled with $N$ fermions. We have also used 
eq.(\ref{cutoff}) to eliminate the cut-off $\Lambda_2$ in favor of $G$ and 
$\Phi_0$.

Since the modes localized on the wall have $\Psibar_0 \Psi_0 = 0$, they have no
effect on the self-consistency of the domain wall solution itself. In 
particular, one can also fill the negative energy states for those modes and 
thus construct the usual vacuum for the 2-d physics on the brane. By 
occupying a localized state $\Psi_0(x_3)$ with spatial momentum $- k_1$, we 
then obtain a left-moving particle with energy $E_0 = - k_1$. At low energies 
$E_0 \ll \Phi_0$ the fermions that are localized on the brane do not feel the
extra dimension and see just a 2-d space-time. At sufficiently large 
energies $E_0 > \Phi_0$, on the other hand, fermions can escape into the extra 
dimension. From the point of view of a 2-d observer, this process seems to 
violate fermion number conservation.

At this point we have explicitly constructed a topologically stable brane on 
which massless left-handed fermions can propagate as free particles. In the 
case of a single domain wall, the dimensional reduction of fermions from three 
to two dimensions is straightforward. In the next Subsection we will construct 
a brane world consisting of a wall-anti-wall pair. In that case, the mechanism 
of dimensional reduction is more subtle.

\subsection{A Brane World Consisting of a Wall-Anti-Wall Pair}

As we have seen, left-handed fermions can be localized on a domain wall.
Similarly, right-handed fermions can be localized on an anti-wall. A world that
contains both left- and right-handed fermions should hence contain a 
wall-anti-wall pair. In order to keep the fermions light, the wall and the
anti-wall must be separated in the extra dimension by a sufficient distance. If
the low-energy physics of the dimensionally reduced theory happens 
simultaneously on the wall and the anti-wall, the question arises how the brane
separation $\beta$ compares with the natural length scale $\xi$ on the 
branes. Interestingly, in analogy to the mechanism of dimensional reduction 
from the symmetric phase that was discussed in Section 4, we will find that, as
$\beta$ increases, the length scale $\xi$ grows exponentially such that the 
brane separation becomes small in the natural physical units of a low-energy 
observer.

Again inspired by the corresponding soliton solution of the 2-d Gross-Neveu
model, we now make the ansatz 
\begin{equation}
\label{profile2}
\Phi(x_3) = \Phi_0 [a \tanh(a \Phi_0 x_3 + c) - a \tanh(a \Phi_0 x_3 - c) - 1].
\end{equation}
As before, we need to solve the Dirac equation in the given background field.
Remarkably, this is still possible in closed form. The following relations
are important for the derivation of the results presented below
\begin{eqnarray}
&&[1 + a \tanh(y - c)][1 - a \tanh(y + c)] = 1 - a^2, 
\nonumber \\
&&\frac{\cosh(y + c)}{\cosh(y - c)} =
\frac{1}{\sqrt{1 - a^2}} [1 + a \tanh(y - c)], \nonumber \\
&&\frac{\cosh(y - c)}{\cosh(y + c)} =
\frac{1}{\sqrt{1 - a^2}} [1 - a \tanh(y + c)], \nonumber \\
&&\frac{a}{\cosh(y + c) \cosh(y - c)} =
\sqrt{1 - a^2}[\tanh(y + c) - \tanh(y - c)]. 
\end{eqnarray}
Here we have put $a \Phi_0 x_3 = y$. The Dirac equation, as well as the above
relations, are satisfied only if
\begin{equation}
\tanh(a \Phi_0 \beta) = \tanh(2 c) = a.
\end{equation}
The parameter $a \in [0,1]$ determines the brane separation $\beta$ and will
later be fixed self-consistently. At zero distance we have $a = 0$ and hence 
$\Phi(x_3) = - \Phi_0$, so that we are in one of the two vacua of the 3-d 
theory. As $a$ approaches 1, on the other hand, $\beta$ goes to infinity. Then 
$\Phi(x_3) = \Phi_0$ and we are in the other vacuum state of the 3-d theory. 
Hence, by varying $a$ we can change the brane separation and interpolate 
smoothly between the two vacua.

In the wall-anti-wall case one obtains the states
\begin{equation}
\Psi_0(x_3) = \frac{\sqrt{a \Phi_0}}{2 \sqrt{E_0(E_0 + k_1)}}
\left(\begin{array}{c} 
- i (E_0 + k_1) \cosh^{-1}(a \Phi_0 x_3 - c) \\ 
m \cosh^{-1}(a \Phi_0 x_3 + c) \end{array} \right),
\end{equation} 
localized on the branes. Here
\begin{equation}
m = \sqrt{1 - a^2} \Phi_0,
\end{equation}
is the mass of the particles propagating on the branes and
\begin{equation}
E_0 = \pm \sqrt{k_1^2 + (1 - a^2) \Phi_0^2} = \pm \sqrt{k_1^2 + m^2},
\end{equation}
is the corresponding energy. These states describe a Dirac fermion of momentum 
$k_1$ whose left-handed component propagates on the wall, while its 
right-handed component propagates on the anti-wall. The wall-anti-wall profile,
as well as the wave function for the left- and right-handed components of a
localized state at rest, are depicted in Figure 6.
\begin{figure}[htb]
\begin{center}
\epsfig{figure=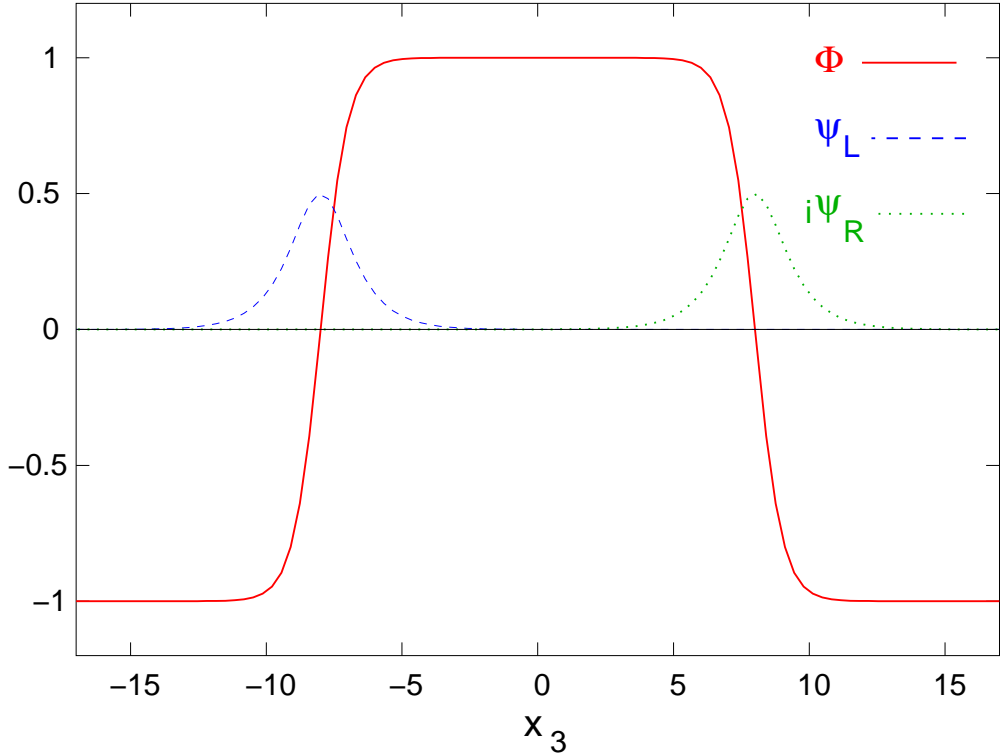,width=10cm,angle=270}
\end{center}
\caption{\it The profile of a wall-anti-wall pair and the left- and 
right-handed components of a localized state at rest with $k_1 = 0$ (for 
$\Phi_0 = 1$, $c = 8$, i.e.\ $a = \tanh(16)$).}
\end{figure}
For a particle moving to the left at high speed one has $E_0 \approx - k_1$. 
Then the upper component of the wave function, which is localized on the 
anti-wall, tends to zero, and the particle is almost entirely located on the 
wall. This is illustrated in Figure 7. 
\begin{figure}[htb]
\begin{center}
\epsfig{figure=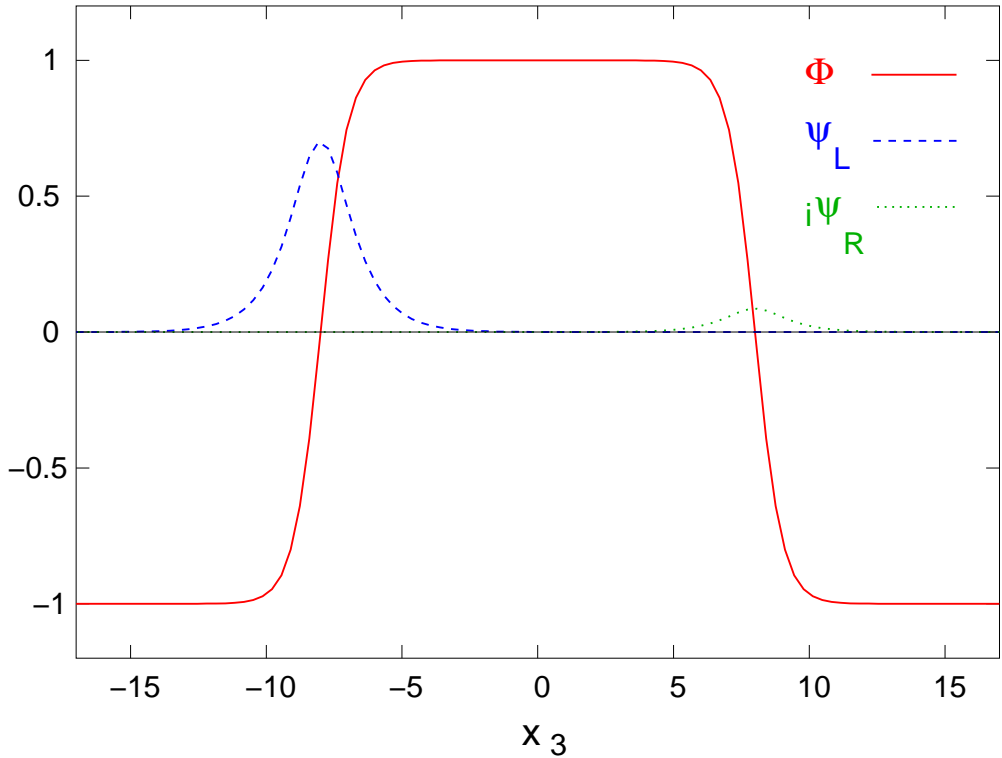,width=10cm,angle=270}
\end{center}
\caption{\it The profile of a wall-anti-wall pair and the left- and 
right-handed components of a left-moving localized state with $k_1 = -4m$ (for 
$\Phi_0 = 1$, $c = 8$, i.e.\ $a = \tanh(16)$).}
\end{figure}
The wave function of a fast right-mover, on the other hand, is concentrated on 
the anti-wall. 

At $a = 0$, i.e.\ at distance $\beta = 0$ between the branes, we end up in the 
vacuum $\Phi = - \Phi_0$ and the mass of the Dirac fermion is simply given by 
the 3-d fermion mass $m = \Phi_0$. When the branes are separated by a large 
distance $\beta \gg 1/\Phi_0$, i.e.\ for $a \rightarrow 1$, on the other hand, 
the mass of the Dirac fermion goes to zero as
\begin{equation}
\label{massgap}
m = 2 \Phi_0 \exp(- \beta \Phi_0).
\end{equation}
Consequently, when the branes are separated at a large distance $\beta$, the 
2-d correlation length $\xi = 1/m \gg \beta$ again grows exponentially with 
$\beta$ and the system dimensionally reduces. In particular, from the point of 
view of a low-energy 2-d observer, the left- and right-handed modes 
propagating simultaneously on the wall and the anti-wall are indistinguishable 
from an ordinary point-like Dirac fermion. At the same time, a 3-d 
high-energy observer whose natural length scale is $1/\Phi_0$ would say that 
the Dirac particle consists of a left-handed and a right-handed constituent, 
both separated by a large distance $\beta \gg 1/\Phi_0$.

In the wall-anti-wall configuration the bulk states take the form
\begin{equation}
\Psi_{k_3}(x_3) = 
\frac{\exp(i k_3 x_3)}{\sqrt{2E(E + k_1)(k_3^2 + a^2 \Phi_0^2)}} 
\left(\begin{array}{c}
i (E + k_1)[a \Phi_0 \tanh(a \Phi_0 x_3 - c) - i k_3] \\
- (\Phi_0 + i k_3)[a \Phi_0 \tanh(a \Phi_0 x_3 + c) - i k_3] 
\end{array} \right),
\end{equation}
with the energy eigenvalues $E = \pm \sqrt{k_1^2 + k_3^2 + \Phi_0^2}$. 

We still need to verify the self-consistency of the wall-anti-wall profile. As
before, we need to sum $\Psibar \Psi$ over all occupied states. First of all, 
we fill all bulk states with negative energy, and we obtain
\begin{eqnarray}
\label{psibarpsi}
&&\Psibar_{k_3}(x_3) \Psi_{k_3}(x_3) = \nonumber \\
&&- \frac{1}{E} \Phi_0 \left(\frac{k_3^2 + \Phi_0^2}{k_3^2 + a^2 \Phi_0^2}
[a \tanh(a \Phi_0 x_3 + c) - a \tanh(a \Phi_0 x_3 - c)] - 1 \right)
= \nonumber \\
&&- \frac{1}{E} \Phi(x_3) - 
\frac{1}{E} \frac{a \Phi_0 m^2}{k_3^2 + a^2 \Phi_0^2}
[\tanh(a \Phi_0 x_3 + c) - \tanh(a \Phi_0 x_3 - c)].
\end{eqnarray}
As in the case of a single wall, the first term, $- \Phi(x_3)/E$, contributes
\begin{equation}
\frac{G}{N} \Psibar(x_3) \Psi(x_3) = 
\Phi(x_3) \frac{G}{2 \pi} \int_0^{\Lambda_2}
\frac{k \, dk}{\sqrt{k^2 + \Phi_0^2}} = \Phi(x_3) 
\frac{G}{2 \pi}(\Lambda_2 - \Phi_0) = \Phi(x_3),
\end{equation}
which is just what we need for self-consistency. However, in the wall-anti-wall
case, there is also the second term. Performing the $k_3$-integration, that
term contributes
\begin{equation}
\label{contribution}
\int_{- \infty}^\infty dk_3 \
\frac{1}{\sqrt{k_1^2 + k_3^2 + \Phi_0^2}}
\frac{a \Phi_0}{k_3^2 + a^2 \Phi_0^2} = \frac{1}{\sqrt{k_1^2 + m^2}}
\left(\pi - 2 \arcsin(\frac{a \Phi_0}{\sqrt{k_1^2 + \Phi_0^2}})\right),
\end{equation}
which needs to be canceled by an appropriate contribution from the states 
localized on the branes. For these states one obtains
\begin{equation}
\Psibar_0(x_3) \Psi_0(x_3) = \frac{1}{E_0} \frac{m^2}{2}
[\tanh(a \Phi_0 x_3 + c) - \tanh(a \Phi_0 x_3 - c)].
\end{equation}
First, let us fill all negative energy states localized on the branes. Those
have energy $E_0 = - \sqrt{k_1^2 + m^2}$ and exactly cancel the contribution 
$\pi$ in eq.(\ref{contribution}). In order to cancel the $\arcsin$-term in 
eq.(\ref{contribution}) as well, we also need to fill some positive energy 
states (or, equivalently, empty some negative energy states). Hence, in 
contrast to the single wall case, the wall-anti-wall configuration is unstable 
if the 2-d brane world is in its vacuum state. In order to stabilize the
brane configuration, we occupy all positive energy states localized on the wall
(with energy $E_0 = \sqrt{k_1^2 + m^2}$) up to some Fermi momentum $k_F$. The
cancellation condition, which determines the value of $k_F$, then takes the form
\begin{equation}
\int_{- k_F}^{k_F} \frac{dk_1}{\sqrt{k_1^2 + m^2}} =
\int_{- \infty}^\infty \frac{dk_1}{\sqrt{k_1^2 + m^2}}
\frac{2}{\pi} \arcsin(\frac{a \Phi_0}{\sqrt{k_1^2 + \Phi_0^2}}) = 
\log\frac{1 + a}{1 - a} \ \Rightarrow \ k_F = a \Phi_0.
\end{equation}
Consequently, the energy of the particles at the Fermi-surface
\begin{equation}
\sqrt{k_F^2 + m^2} = \sqrt{a^2 \Phi_0^2 + (1 - a^2) \Phi_0^2} = \Phi_0,
\end{equation}
is equal to the lowest energy $\Phi_0$ of states propagating in the 3-d
bulk of the extra dimension. Any fermion that one adds to the brane world has
enough energy to escape into the extra dimension. Thus, our brane world --- 
which indeed contains naturally light fermions --- is necessarily completely 
packed with these particles. Hence, any potentially interesting physics of the 
light fermions is entirely Pauli-blocked, and this ``world'' can only exist in
one (physically quite uninteresting) state. This is in contrast to the 
single-wall case, where any fermion occupation of the localized states was 
self-consistent with the wall profile. It should be noted that we have assumed 
the configuration to be constant in the spatial 1-direction along the brane. 
The results of Ref.\ \cite{Thi03} show that, at non-zero fermion density, 
translation invariance may break by the spontaneous formation of a crystal 
lattice. If this happens here as well, it is still possible that our brane 
world displays some non-trivial physics. This could be investigated along the 
lines of Ref.\ \cite{Thi03}.

\subsection{Chiral Order Parameter of the 2-d Theory}

As we have seen, in the wall-anti-wall case a light fermion mode is localized
on the branes. In this Subsection we raise the question about the dynamical 
origin of the fermion mass. In particular, we ask if it arises as a consequence
of chiral symmetry breaking in the dimensionally reduced theory. 

Let us first discuss the single wall case. Obviously, the wall configuration 
arises as a consequence of the broken $\Z(2)$ symmetry in the 3-d bulk. This
$\Z(2)$ symmetry is a combination of a 2-d chiral transformation and a 
reflection about a 2-d plane perpendicular to the 3-direction. It is 
interesting to note that the wall profile $\Phi(x_3) = \Phi_0 
\tanh(\Phi_0 x_3)$ is invariant against the transformation
\begin{equation}
\label{trans}
\Phi(x_1,x_2,x_3)' = - \Phi(x_1,x_2,- x_3),
\end{equation}
which combines a 2-d chiral transformation with a reflection at the domain wall
center. Together with the transformation of eq.(\ref{symmetry3}), this 
particular $\Z(2)$ symmetry indeed remains intact even in the presence of the 
domain wall. From the point of view of a 2-d observer living on the brane, this
$\Z(2)$ symmetry is nothing but 2-d chiral symmetry. As a consequence of this
unbroken $\Z(2)$ chiral symmetry, the fermion localized on the brane is 
massless. 

Next we discuss the wall-anti-wall case in which the wall profile is no longer
invariant under the transformation of eq.(\ref{trans}). As a consequence, there
is no exact $\Z(2)$ chiral symmetry from the point of view of the 2-d brane
world. Still, as the wall and the anti-wall are separated by a large distance
$\beta$, an approximate chiral symmetry emerges which turns into an exact 
symmetry at infinite brane separation. The fermion mass 
$m = 2 \Phi_0 \exp(- \beta \Phi_0)$ decreases exponentially with $\beta$. Is 
this mass due to explicit or spontaneous breaking of the emerging approximate
$\Z(2)$ symmetry? In order to clarify this question, we now compute the value
of the chiral order parameter of the dimensionally reduced 2-d theory on the
brane. First, we define an effective 2-d fermion field
\begin{eqnarray}
&&\psi_R(x_1,x_2) = \int dx_3 \ \Psi_R(x_1,x_2,x_3) \sqrt{\frac{a \Phi_0}{2}}
\cosh^{-1}(a \Phi_0 x_3 - c), \nonumber \\
&&\psi_L(x_1,x_2) = \int dx_3 \ \Psi_L(x_1,x_2,x_3) \sqrt{\frac{a \Phi_0}{2}}
\cosh^{-1}(a \Phi_0 x_3 + c),
\end{eqnarray}
by smearing the 3-d fermion field with the wave function of the modes localized
on the right or left side of the brane world. Note that the 2-d fermion field, 
defined in this way, indeed has the correct dimension. 

Next, we calculate the 2-d chiral condensate $\psibar \psi$. It is 
straightforward to show that only the modes localized on the branes contribute 
to this quantity. The contribution takes the form
\begin{equation}
\psibar_0 \psi_0 = \frac{m}{E_0},
\end{equation}
where again $E_0 = \pm \sqrt{k_1^2 + m^2}$ and 
$m = \sqrt{1 - a^2} \, \Phi_0$. The
domain wall height $\Phi_0$ serves as a natural cut-off for the physics in the
brane world. Hence, we occupy all states with negative energies 
$E_0 \geq - \Phi_0$, and (for $\Phi_0 \gg m$) we obtain
\begin{equation}
\phi_0 = \frac{g}{N} \psibar \psi = 
- m \frac{g}{2 \pi} \int_{- \Phi_0}^{\Phi_0} \frac{dk_1}{\sqrt{k_1^2 + m^2}} = 
- \frac{g m}{\pi} \log(\frac{2 \Phi_0}{m}).
\end{equation}
Using eq.(\ref{massgap}) we identify 
\begin{equation}
\frac{1}{g} = \frac{\beta \Phi_0}{\pi} = 
2 \beta \left(\frac{1}{G_c} - \frac{1}{G}\right),
\end{equation}
as the effective 2-d coupling constant, and we then indeed obtain 
\begin{equation}
m = |\phi_0|,
\end{equation} 
just as in the 2-d Gross-Neveu model. As we have seen in the previous 
Subsection, for reasons of consistency, our brane world cannot exist in its 
vacuum state. Thus, the actual (non-vacuum) value of the chiral condensate also
receives contributions from states with positive energies $E_0 \leq \Phi_0$ 
and, in fact, even vanishes. Still, since for large $\beta$ the vacuum value of
the chiral condensate $\phi_0$ agrees with the dynamically generated fermion 
mass $m$, we conclude that this mass actually results from the spontaneous 
breakdown of the emergent $\Z(2)$ chiral symmetry.

\subsection{Brane Tension and Brane World Stability}

Let us consider the question of stability of the brane world. Obviously, unlike
the single wall, the wall-anti-wall configuration is not topologically stable. 
For example, one might worry that the wall and the anti-wall attract each 
other and annihilate through a continuous deformation of the profile 
$\Phi(x_3)$ into the trivial vacuum $\Phi(x) = - \Phi_0$. The energy stored in 
the brane tension would then be released and turned into heavy 3-d fermions. 
Such a catastrophic event would clearly be the end of our brane world. 
Alternatively, if the wall and the anti-wall repel each other, they would 
simply drift apart in the extra dimension. Remarkably, in the Gross-Neveu model
the wall-anti-wall configuration is stable against both annihilation and 
drifting apart. What is the dynamical mechanism responsible for the stability? 
The key observation is that the wall and the anti-wall themselves consist of 
fermions --- what else could they possibly be made of in this model? 

In the wall-anti-wall case, the fermion density is given by
\begin{equation}
\frac{F}{L} = \frac{N}{2 \pi} \int_{- k_F}^{k_F} dk_1 = \frac{N a \Phi_0}{\pi}.
\end{equation}
Let us also calculate the brane tension $\sigma = E/L$, i.e.\ its energy per 
unit length. The brane tension receives contributions from three different 
sources: the filled bulk states, the filled surface states localized on the 
branes, and the $\Phi^2$ term. The latter contributes
\begin{equation}
\sigma_1 = \frac{N}{2G} \int_{- \infty}^\infty dx_3 \ [\Phi(x_3)^2 - \Phi_0^2]
= - \frac{2 N a \Phi_0}{G},
\end{equation}
while the surface states localized on the branes yield
\begin{equation}
\sigma_2 = - \frac{N}{2 \pi} \int dk_1 \ \sqrt{k_1^2 + m^2} +
\frac{N}{2 \pi} \int_{-k_F}^{k_F} dk_1 \ \sqrt{k_1^2 + m^2}.
\end{equation}
The calculation of the fermionic contribution of the bulk states requires
special care. Following Refs.\ \cite{Pau91,Gra01}, we impose periodic boundary 
conditions in the 3-direction over a finite extent $L$. Note that, in contrast 
to a single wall, the wall-anti-wall configuration is indeed consistent with 
periodic boundary conditions. In the finite box only discrete values for 
$k_{3,n}$ are allowed. These values are determined by the scattering phase 
shift $\delta$ as
\begin{equation}
k_{3,n} L + \delta(k_{3,n}) = 2 \pi n,
\end{equation}
with $n \in \Z$. For sufficiently large $L$ (such that 
$\tanh(a \Phi_0 L/2 + c) \approx 1$) the $k_{3,n}$ values are given by
\begin{equation}
\exp(i k_{3,n} L) =\exp(- i \delta(k_{3,n})) = 
\frac{i k_{3,n} + a \Phi_0}{i k_{3,n} - a \Phi_0}.
\end{equation}
In the next step we sum the energy differences between occupied bulk states in 
the wall-anti-wall and the vacuum configuration
\begin{eqnarray}
\sigma_3&=&
- \frac{N}{2 \pi} \int dk_1 \sum_n \left[\sqrt{k_1^2 + k_{3,n}^2 + \Phi_0^2} -
\sqrt{k_1^2 + (2 \pi n/L)^2 + \Phi_0^2}\right] \nonumber \\
&=&- \frac{N}{2 \pi} \int dk_1 \sum_n \left(k_{3,n} - \frac{2 \pi n}{L}\right) 
\frac{d\sqrt{k_1^2 + k_3^2 + \Phi_0^2}}{dk_3}|_{k_3 = 2 \pi n/L} \nonumber \\
&=&\frac{N}{2 \pi} \int dk_1 \ \frac{1}{L} \sum_n \delta(\frac{2 \pi n}{L}) 
\frac{d\sqrt{k_1^2 + k_3^2 + \Phi_0^2}}{dk_3}|_{k_3 = 2 \pi n/L} \nonumber \\
&=&\frac{N}{(2 \pi)^2} \int dk_1 \int dk_3 \ \delta(k_3)
\frac{d\sqrt{k_1^2 + k_3^2 + \Phi_0^2}}{dk_3}.
\end{eqnarray}
In these manipulations, we have Taylor expanded around $k_3 = 2 \pi n/L$ and, 
at the end, we have taken the $L \rightarrow \infty$ limit. Performing a 
partial integration, using
\begin{equation}
\frac{d\delta(k_3)}{dk_3} = - \frac{2 a \Phi_0}{k_3^2 + a^2 \Phi_0^2},
\end{equation}
and summing up all contributions to the brane tension, one finally obtains 
\begin{equation}
\frac{E}{L} = \sigma = \sigma_1 + \sigma_2 + \sigma_3 = 
\frac{N a \Phi_0^2}{\pi}.
\end{equation}
Remarkably, the energy per fermion, $E/F = \Phi_0$, is independent of the
separation of the walls and equals the mass of a fermion in the 3-d bulk.
Consequently, given a fixed number of fermions, they can be divided arbitrarily
into some that form the brane world and others that remain at rest in the 3-d
bulk. This explains how our brane world is protected from the dooms-day 
scenario of wall-anti-wall annihilation. The branes can accrete fermions that 
fall onto them, coming from the bulk of the extra dimension. This process 
increases the brane separation $\beta$ as well as the Fermi-momentum $k_F$, but
it also decreases the fermion mass $m$, such that the energy of the states at 
the Fermi-surface $\Phi_0$ remains fixed. The thickness of the world $\beta$ 
increases only logarithmically with the fermion density. 

At infinite brane separation, $\beta \rightarrow \infty$, i.e.\ for 
$a \rightarrow 1$, the fermion density and the energy density approach two 
times the corresponding values of a single wall brane world that is completely 
packed with fermions. The brane tension for this object is given by
$N \Phi_0^2/2 \pi$. Interestingly, once the wall and the anti-wall are 
separated by an infinite distance, they can support configurations with any 
value of the Fermi-momentum $k_F \leq \Phi_0$. The corresponding brane tension
is then given by
\begin{equation}
\sigma(k_F) = \frac{N}{4 \pi}(\Phi_0^2 + k_F^2).
\end{equation}
For instance, an empty single-wall brane world (in its vacuum state) costs an 
energy $\sigma(0) = N \Phi_0^2/4 \pi$ per unit length.

\section{Conclusions}

As we discussed in the Introduction, light fermions can arise naturally from
higher dimensions as states localized on domain walls. Following the ideas 
behind D-theory, in contrast to standard applications of domain wall fermions 
in lattice field theory, we have taken the extra dimension physically 
seriously. In particular, we have maintained locality in the bulk of the extra 
dimension, and we have used dimensional reduction to make the extra dimension 
invisible to a low-energy observer. The 3-d Gross-Neveu model at large $N$ has 
been used as a toy model in which one can obtain analytical insight into these
phenomena. In particular, the Gross-Neveu model with a discrete $\Z(2)$ chiral 
symmetry provides domain walls dynamically through spontaneous chiral symmetry 
breaking. Interestingly, fermionic zero-modes can be localized on these walls 
which themselves consist of fermions. An exact analytic solution was found for 
a stable wall-anti-wall configuration, and indeed light fermions arise 
naturally via dimensional reduction in this brane world. Remarkably, the 
wall-anti-wall configuration is stable (although not topologically stable) due 
to the baryon asymmetry of the fermions that the brane is made of. This 
mechanism of brane stabilization may be interesting for realistic brane world 
constructions. Ironically, in our toy model the ``world'' is stable only if the
baryon asymmetry is so large that all fermion states are occupied and all 
physics is completely Pauli-blocked. We take this as a lesson for brane world 
model building. Our toy ``world'' was free to follow its own dynamics, without 
us making any ad-hoc assumptions about where branes should be located. Perhaps 
not surprisingly, the ``world'' then behaved in some --- but not in all 
respects --- as its builders had intended.

\section*{Acknowledgements}

We would like to thank S.~Chandrasekharan, N.~Graham, P.~Hasenfratz, and 
R.~L.\ Jaffe for very interesting discussions. W.B. thanks the
Universit\"{a}t Bern for kind hospitality during several visits.
This work was supported in part
by funds provided by the Schweizerischer Nationalfond (SNF), by the European 
Community's Human Potential Programme under contract HPRN-CT-2000-00145, and by
the DFG Sonderforschungsbereich Transregio 9, ``Computergest\"utzte 
Theoretische Teilchenphysik''.


\begin{thebibliography}{10}

\bibitem{Wil02}
F.~Wilczek, hep-ph/0201222.

\bibitem{Cal85}
C.~G.~Callan and J.~A.~Harvey, Nucl.\ Phys.\ B250 (1985) 427.

\bibitem{Kap92}
D.~B.~Kaplan, Phys.\ Lett.\ B288 (1992) 342.

\bibitem{Neu93}
H.~Neuberger and R.~Narayanan, Phys.\ Rev.\ Lett.\ 71 (1993) 3251; 
Nucl.\ Phys.\ B412 (1994) 574.

\bibitem{Gin82}
P.~H.~Ginsparg and K.~G.~Wilson, Phys.\ Rev.\ D25 (1982) 2649.

\bibitem{Lue98}
M.~L\"uscher, Phys.\ Lett.\ B428 (1998) 342.

\bibitem{Lue99}
M.~L\"uscher, Nucl.\ Phys.\ B549 (1999) 295.

\bibitem{Wie93}
U.-J.~Wiese, Phys.\ Lett.\ B315 (1993) 417.

\bibitem{Bie96}
W.~Bietenholz and U.-J.~Wiese, Nucl.\ Phys.\ B464 (1996) 319.

\bibitem{DeG97}
T.~DeGrand, A.~Hasenfratz, P.~Hasenfratz, P.~Kunszt, and
F.~Niedermayer, Nucl.\ Phys.\ B (Proc.\ Suppl.) 53 (1997) 942.

\bibitem{Has98}
P.~Hasenfratz, Nucl.\ Phys.\ B (Proc.\ Suppl.) 63 (1998) 53.\\
P.~Hasenfratz, V.~Laliena, and F.~Niedermayer, Phys.\ Lett.\ B427 (1998) 125.

\bibitem{Ark98}
N.~Arkani-Hamed, S.~Dimopoulos, and G.~Dvali, Phys.\ Lett.\ B429 (1998) 263.

\bibitem{Ran99}
L.~Randall and R.~Sundrum, Phys.\ Rev.\ Lett.\ 83 (1999) 3370; 4690.

\bibitem{Cha97}
S.~Chandrasekharan and U.-J.~Wiese, Nucl.\ Phys.\ B492 (1997) 455; \\
R.~C.~Brower, S.~Chandrasekharan, and U.-J.~Wiese, Phys.\ Rev.\ D60 (1999) 
094502; \\
U.-J.~Wiese, Nucl.\ Phys.\ B (Proc.\ Suppl.) 73 (1999) 146.

\bibitem{Gro74}
D.~J.~Gross and A.~Neveu, Phys.\ Rev.\ D10 (1974) 3235.

\bibitem{Izu00}
T.~Izubuchi and K.-I.~Nagai, Phys.\ Rev.\ D61 (2000) 094501.

\bibitem{Lat03} W.~Bietenholz, G.~Gfeller, and U.-J.~Wiese,
hep-lat/0308022.

\bibitem{Sch01}
V.~Sch\"on and M.~Thies, in ``At the frontiers of particle physics: Handbook of
QCD'', vol.\ 3, p.\ 1945, ed.\ M.~Shifman, World Scientific (2001).

\bibitem{Gat92}
G.~Gat, A.~Kovner, and B.~Rosenstein, Nucl.\ Phys.\ B385 (1992) 76.

\bibitem{Ana98}
G.~N.~J.~Ananos, A.~P.~C.~Malbouisson, M.~B.~Silva-Neto, and N.~F.~Svaiter, 
Physica A260 (1998) 157.

\bibitem{Hof02}
F.~H\"ofling, C.~Nowak, and C.~Wetterich, Phys.\ Rev.\ B66 (2002) 205111.

\bibitem{Cha88}
S.~Chakravarty, D.~R.~Nelson, and B.~I.~Halperin, Phys.\ Rev.\ Lett.\ 60 (1988)
1057; Phys.\ Rev.\ B39 (1989) 2344.

\bibitem{Has91}
P.~Hasenfratz and F.~Niedermayer, Phys.\ Lett.\ B268 (1991) 231.

\bibitem{Das75}
F.~Dashen, B.~Hasslacher, and A.~Neveu, Phys.\ Rev.\ D11 (1975) 3424;
Phys.\ Rev.\ D12 (1975) 2443.

\bibitem{Pau91}
R.~Pausch, M.~Thies, and V.~L.~Dolman, Z.\ Phys.\ A338 (1991) 441.

\bibitem{Fei95}
J.~Feinberg, Phys.\ Rev.\ D51 (1995) 4503; hep-th/0305240.

\bibitem{Cha94}
S.~Chandrasekharan, Phys.\ Rev.\ D49 (1994) 1980.

\bibitem{Thi03}
M.~Thies and K.~Urlichs, Phys.\ Rev.\ D67 (2003) 125015; \\
M.~Thies, hep-th/0308164.

\bibitem{Gra01}
N.~Graham, R.~L.~Jaffe, M.~Quandt, and H.~Weigel, Phys.\ Rev.\ Lett.\ 87 (2001)
131601.

\end{thebibliography}
\end{document}